\documentclass[prb,twocolumn,floatfix,preprintnumbers,amsmath,amssymb,superscriptaddress]{revtex4}

\usepackage{graphicx}
\usepackage{dcolumn}
\usepackage{bm}
\usepackage{color}
\usepackage{xspace}
\usepackage{float}
\usepackage{soul}

\begin{document}

\title{Raman study of spin excitations in the tunable quantum spin ladder Cu(Qnx)(Cl$_{1-x}$Br$_x$)$_2$}

\author{G. Simutis}
 \email{gsimutis@phys.ethz.ch}
\affiliation{Neutron Scattering and Magnetism, Laboratory for Solid State
Physics, ETH Z\"urich, Z\"urich, Switzerland}

\author{S. Gvasaliya}
\affiliation{Neutron Scattering and Magnetism, Laboratory for Solid State
Physics, ETH Z\"urich, Z\"urich, Switzerland}

\author{F.Xiao}
\affiliation{Department of Physics, Durham University, Durham, United Kingdom}

\author{C. P. Landee}
\affiliation{Department of Physics, Clark University, Worcester, MA 01610, USA}

\author{A. Zheludev}
\affiliation{Neutron Scattering and Magnetism, Laboratory for Solid State
Physics, ETH Z\"urich, Z\"urich, Switzerland}

\date{\today}

\begin{abstract}
Raman spectroscopy is used to study magnetic excitations in the quasi one dimensional $S=1/2$ quantum spin systems Cu(Qnx)(Cl$_{1-x}$Br$_x$)$_2$.  The low energy spectrum is found to be dominated by a two-magnon continuum as expected from the numerical calculations for the Heisenberg spin ladder model. The continuum shifts to higher energies as more Br is introduced. The cutoff of the scattering increases faster than the onset indicating that the increase of exchange constant along the leg is the main effect on the magnetic properties. The upper and lower continuum thresholds are measured as a function of Br content across the entire range and compared to estimates based on previous bulk studies. We observe small systematic deviations that are discussed.
\end{abstract}

 \maketitle
\section{Introduction}

Heisenberg $S=1/2$ quantum spin ladders are among the most fundamental models in low dimensional solid state physics.\cite{Rice1993,Dagotto1996,Giamarchibook} The limiting cases of non-interacting dimers and non-interacting spin chains have exact solutions. The ground state and excitations change dramatically with a changing ratio of exchange constants along the ladder leg (J$_{||}$) and rung (J$_\bot$), respectively. On the experimental side, recent progress in organic quantum materials led to the discovery of several excellent spin-ladder prototype compounds. Perhaps the ``cleanest'' realizations of the model are DIMPY - (C$_7$H$_{10}$N)$_2$CuBr$_4$\cite{Shapiro2007,Hong2010,Schmidiger2011} and BPCB -  (C$_5$H$_{12}$N)$_2$CuBr$_4$\cite{Patyal1990,Watson2001,Klanjsek2008,Thielemann2009,Savici2009} for the strong-leg and strong-rung case, respectively. The recently characterized family of materials Cu(Qnx)(Cl$_{1-x}$Br$_x$)$_2$,\cite{Lindroos1990,Landee2003,Keith2011,Hong2006,Somoza2012,Povarov2014} where Qnx stands for quinoxaline (C$_8$H$_6$N$_2$), opened a new opportunity of continuously varying the leg to rung exchange ratio, albeit in a rather narrow range. Most previous studies of these materials concentrated on the bulk measurements and provided rough estimates of these parameters: $J_{||}=1.61$~meV, $J_\bot=2.95$~meV ($J_\bot/J_{||}=1.83$) and $J_{||}=1.99$~meV, $J_\bot=3.26$~meV ($J_\bot/J_{||}=1.64$) for $x=0$ and $x=1$, respectively. To date, the only direct study of the excitation spectrum are neutron experiments on powder samples of the Br-rich material\cite{Hong2006}. Unfortunately, more detailed neutron studies are hindered by the challenges of growing suitable single crystals and by the need to fully deuterate the organic ligand.

\begin{figure}
\centering
\includegraphics[width={\columnwidth}]{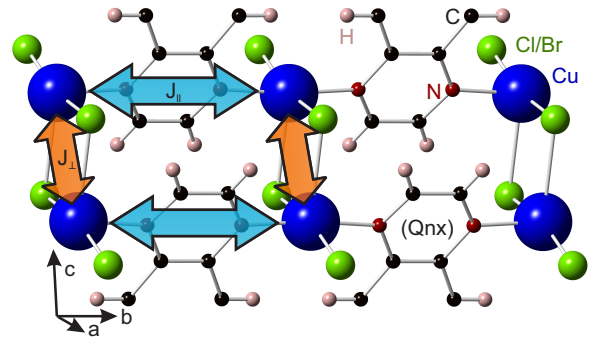}
\caption{Sketch of the Cu(Qnx)(Cl$_{1-x}$Br$_x$)$_2$ ladder with thick arrows showing the principle exchange paths. The interaction along the leg of the ladder proceeds through quinoxaline(Qnx) molecules. Part of H and C ions that are further away from the magnetic ions are not shown. The exchange along the rung is due to halogen ions.\label{fig:structure}}
\end{figure}

An alternative approach to studying magnetic excitations in quantum spin systems is provided by light spectroscopy.\cite{Lemmens2003} This technique has been applied to a variety of materials, including dimer compounds\cite{Choi2003} and spin chains.\cite{Gnezdilov2012,Choi2006} Raman spectroscopy has also proven useful in probing quantum criticality.\cite{Gros2003,Kuroe2012}  For spin ladders, Raman techniques are particularly appealing. The simplest case of 2-magnon scattering essentially probes energy-energy correlations between exchange bonds.\cite{Freitas2000} In the general case, the measured signal is a combination of specific spin spin correlation functions with unknown coefficients that depend on the electronic structure. For the simple Heisenberg spin ladder model though, contributions of interactions on the rungs and legs are proportionate, and therefore the spectrum can be calculated to within a single scale factor.\cite{Schmidt2001,Schmidt2005} Several spin ladder materials have been studied with Raman scattering to date, including BiCu${_2}$PO$_6$\cite{Choi2013}, Sr$_{14-x-y}$Ca$_x$Y$_y$Cu$_{24}$O$_{41}$\cite{Sugai1999,Gozar2001} and  SrCu$_2$O$_3$.\cite{Grossling2003} However, none of those actually correspond to the simple Heisenberg ladder model, involving next nearest neighbor, inter-lader or 4-spin ``cyclic'' exchange interactions.  While some Raman data do exist for the almost perfect BPCB ladder system\cite{Choi2005}, the limited signal strength precludes quantitative analysis.

In the present work we use Raman spectroscopy to investigate magnetic excitations in Cu(Qnx)(Cl$_{1-x}$Br$_x$)$_2$ for the entire range of Br concentrations. We are able to measure the evolutions of both the gap energy and the magnon bandwidth, and study these parameters as a function of Br concentration. Comparing our results to estimates based on previous bulk measurements, we note systematic discrepancies. These findings are discussed in the context of existing numerical calculations of Raman spectra for the ideal Heisenberg spin ladder model.

\section{Experiment}

The Cu(Qnx)(Cl$_{1-x}$Br$_x$)$_2$ crystals used in this study were grown by slow diffusion in methanol solution \cite{Keith2011}. They crystalize in a monoclinic C 2/m space group, with lattice parameters a=13.237 Å, b=6.935 \AA, c=9.775 \AA, $\beta$=107.88$^{\circ}$ for pure Cu(Qnx)Cl$_2$\cite{Lindroos1990} and a=13.175 \AA, b=6.929 \AA, c=10.356 \AA, $\beta$=107.70$^{\circ}$ for pure Cu(Qnx)Br$_2$.\cite{Landee2003} Antiferromagnetic chains of $S=1/2$ Cu$^{2+}$ ions bridged by Qnx molecules run along the crystalographic $b$ axis. As shown in Fig.~\ref{fig:structure}, these chains are coupled into ladders by rung superexchange paths via pairs of halogen ions. As Cl is gradually replaced by Br, the most significant crystalographic change is the increase of the lattice constant $c$, which affects the Cu-Cl/Br-Cu bond angle, and thereby results in a change of the rung exchange constant.\cite{Povarov2014} Interestingly, the magnetic properties are modified more strongly along the leg. This is because of a complex orbital overlap through Qnx molecules.\cite{Somoza2012} A subtle crystalographic change along the leg due to chemical pressure leads to a considerable change in the magnetic properties. As-grown crystals of a typical size 2x1x1 mm$^{3}$ were aligned using an X-Ray diffractometer. In all cases, the surfaces studied contained a well-defined $b$ axis, which coincides with the leg of the ladder.

\begin{figure}
\centering
\includegraphics[width={\columnwidth}]{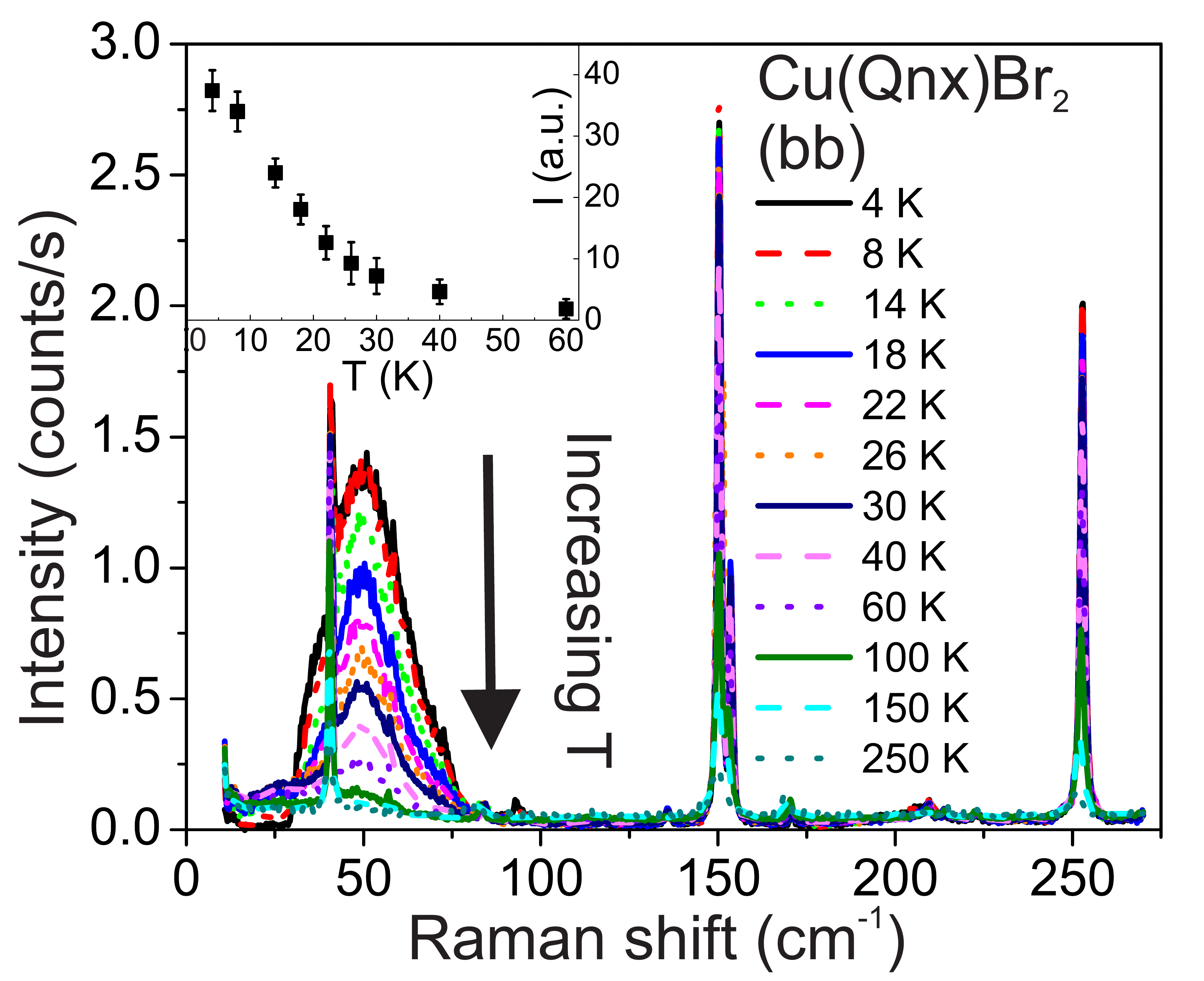}
\caption{Temperature dependence of the spectrum for the Br-end compound. A broad continuum appearing at low temperatures is interpreted as two-magnon scattering. The inset shows the integrated intensity of the scattering continuum as a function of temperature.\label{fig:CQBT}}
\end{figure}

\begin{figure}
\centering
\includegraphics[width={\columnwidth}]{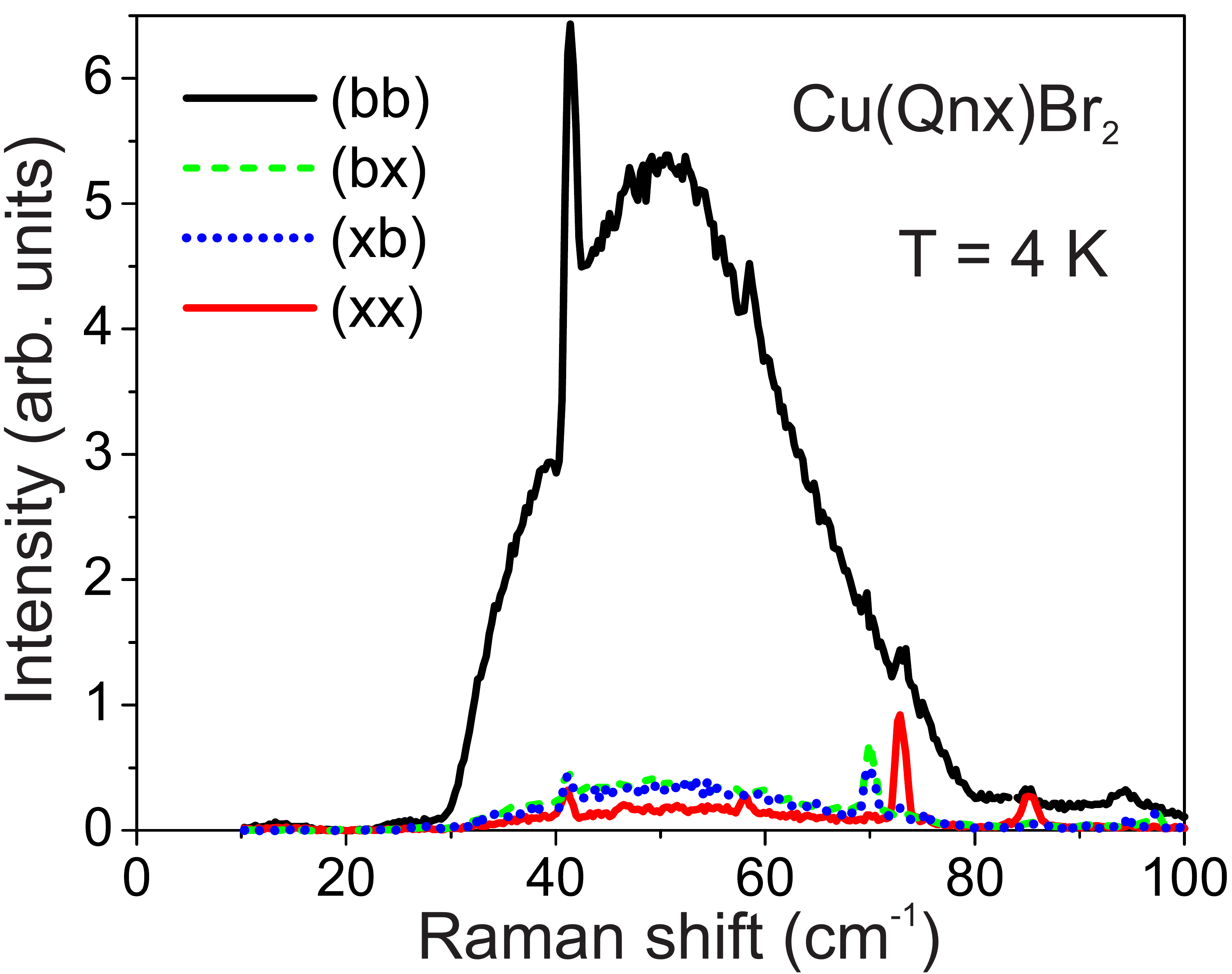}
\caption{Spectra of the Br-end compound in different polarization settings. The strongest magnetic signal is observed in the configuration where polarisation is parallel to the leg of the ladder as expected from theoretical prediction\cite{Schmidt2001}.\label{fig:CQBP}}
\end{figure}

\begin{figure}
\centering
\includegraphics[width={\columnwidth}]{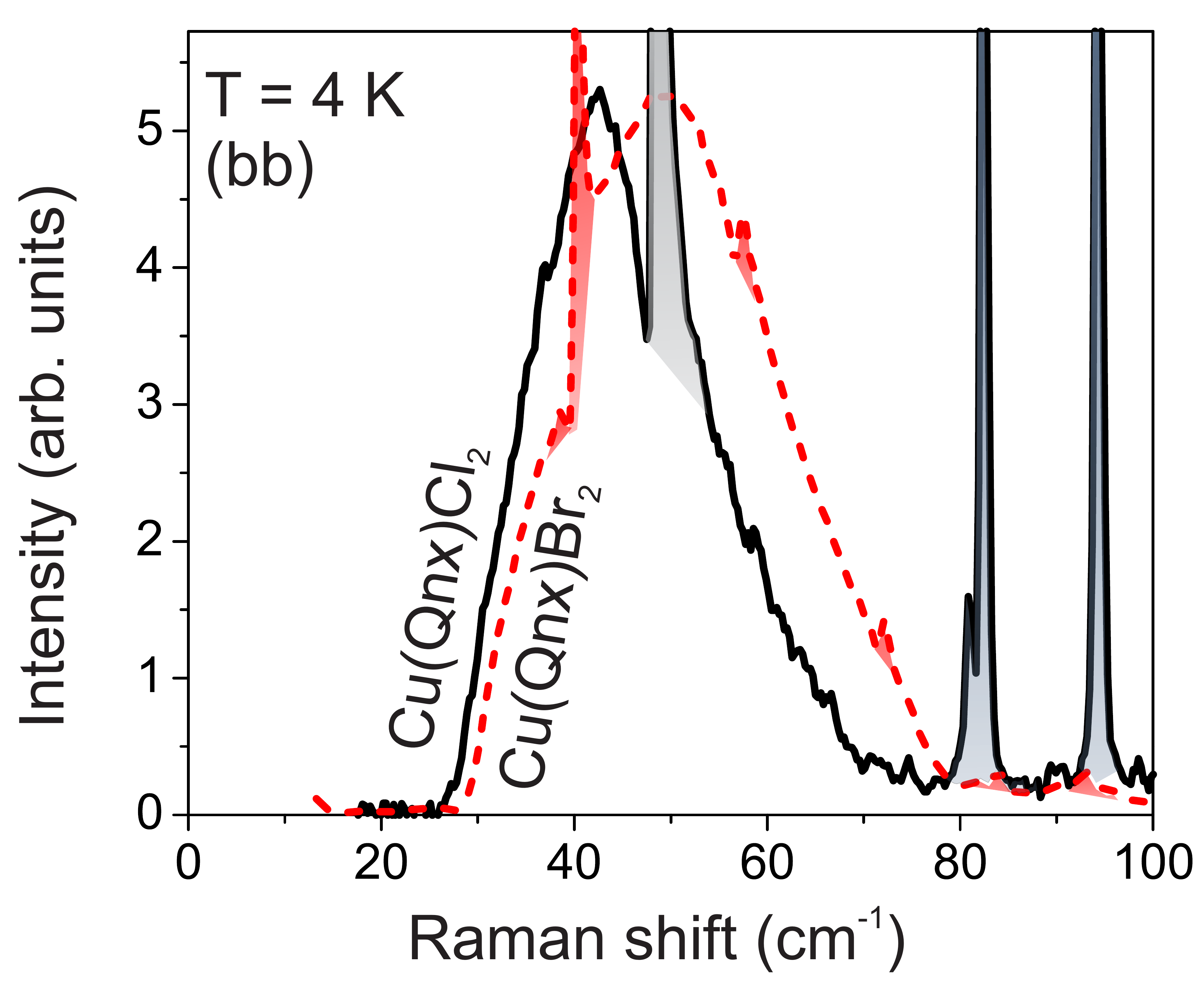}
\caption{The spectra from the two end compounds. The sharp features are due to lattice vibrations. The broad continuum is due to two-magnon scattering. The most intense phonons are shaded.\label{fig:CQBC}}
\end{figure}

Raman spectroscopy measurements were carried out in a backscattering geometry, using a $\lambda$ = 532~nm solid state laser. A low power of 0.5~mW was used in order to limit sample heating. The spectra were obtained a using Trivista tripple spectrometer and a liquid nitrogen cooled CCD detector with reduced etaloning. The samples were cooled down using a helium flow optical cryostat from Cryovac. Most measurements were performed at a temperature of 4~K. In order to obtain good quality spectra at base temperature, measurements were taken in several acquisitions which allowed increasing signal to noise ratio and removing detector response to cosmic muons. The total counting time was of the order of 10 hours per sample for the base temperature measurements. Shorter scans were performed when studying temperature and polarisation dependencies.

The polarisation of incoming and detected light was selected by using a $\lambda$/2 plate and a polariser, respectively. Additionally,  a $\lambda$/4 plate was introduced after the polariser to correct for the efficiency of the grating as function of the polarization of the light. The spectra were normalised to a standard light source, in order to remove the nonuniform response of the spectrometer. In the text below we define the polarizations of incoming and analyzed light in terms of the crystal axes. In this notation,  the polarization of both incoming and analyzed light along the ladder direction (crystallographic $b$ axis) is denoted as `` ($bb$)''. Due to the way the crystals grew, we could not perfectly control the geometry of the experiment in the transverse direction, therefore we use ``(xx)'' to indicate that both polarizations are transverse to the direction of the leg.

\section{Experimental Results}
At high temperatures, the obtained spectra are dominated by phonons. As the temperature is lowered down to about 60 K, a broad continuum of excitations develops at low energies, as shown in Fig.~\ref{fig:CQBT}. From its temperature behavior, comparison with neutron results\cite{Hong2006} and theoretical expectations\cite{Schmidt2001}, we interpret this broad low-energy feature as coming from two-magnon magnetic scattering. The inset shows the integrated intensity of the two-magnon scattering as a function of temperature, which is similar to one observed in a similar strong-rung ladder BPCB.\cite{Choi2005} The shape of this signal remains unchanged in different polarization configurations that we could access, but its intensity is strongest in ($bb$) geometry as shown in Fig. ~\ref{fig:CQBP}. Due to the shape of the crystals we could not access ($yy$) geometry in a controlled manner. From now on all the data shown and discussed here were taken in ($bb$) configuration. The low-energy spectra for the two end compounds in the series are shown in greater detail in Fig.~\ref{fig:CQBC}. The sharp peak in the middle of the magnetic continuum is a phonon related to the halogen ion movement, as it shifts significantly as a function of Br~content.

\begin{figure}
\centering
\includegraphics[width={\columnwidth}]{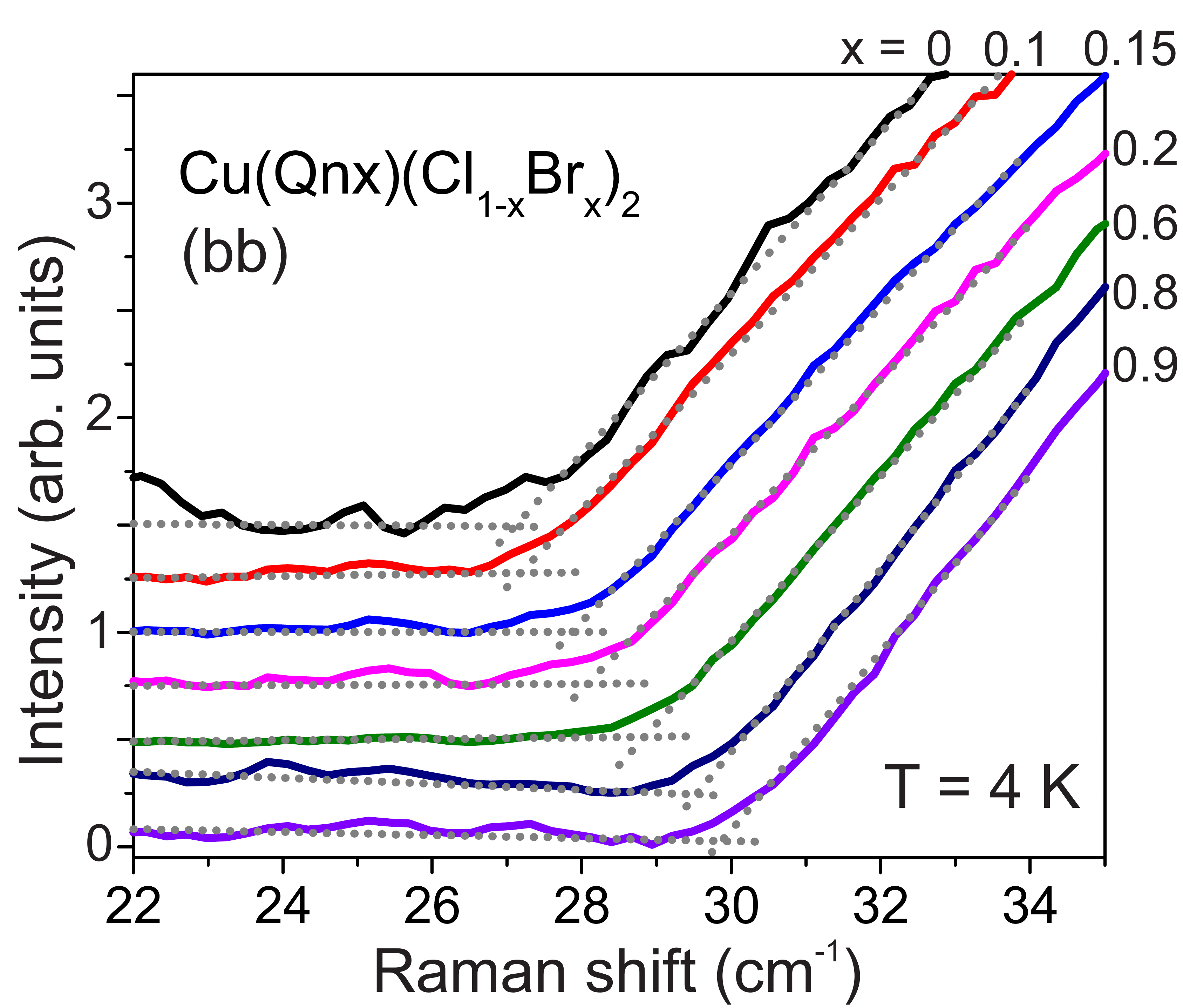}
\caption{The onset region for selected different concentrations.The spectra are offset for a clearer presentation of data. The fitted straight lines are presented as dotted lines showing the estimates of the onset.\label{fig:onset}}
\end{figure}

\begin{figure}
\centering
\includegraphics[width={\columnwidth}]{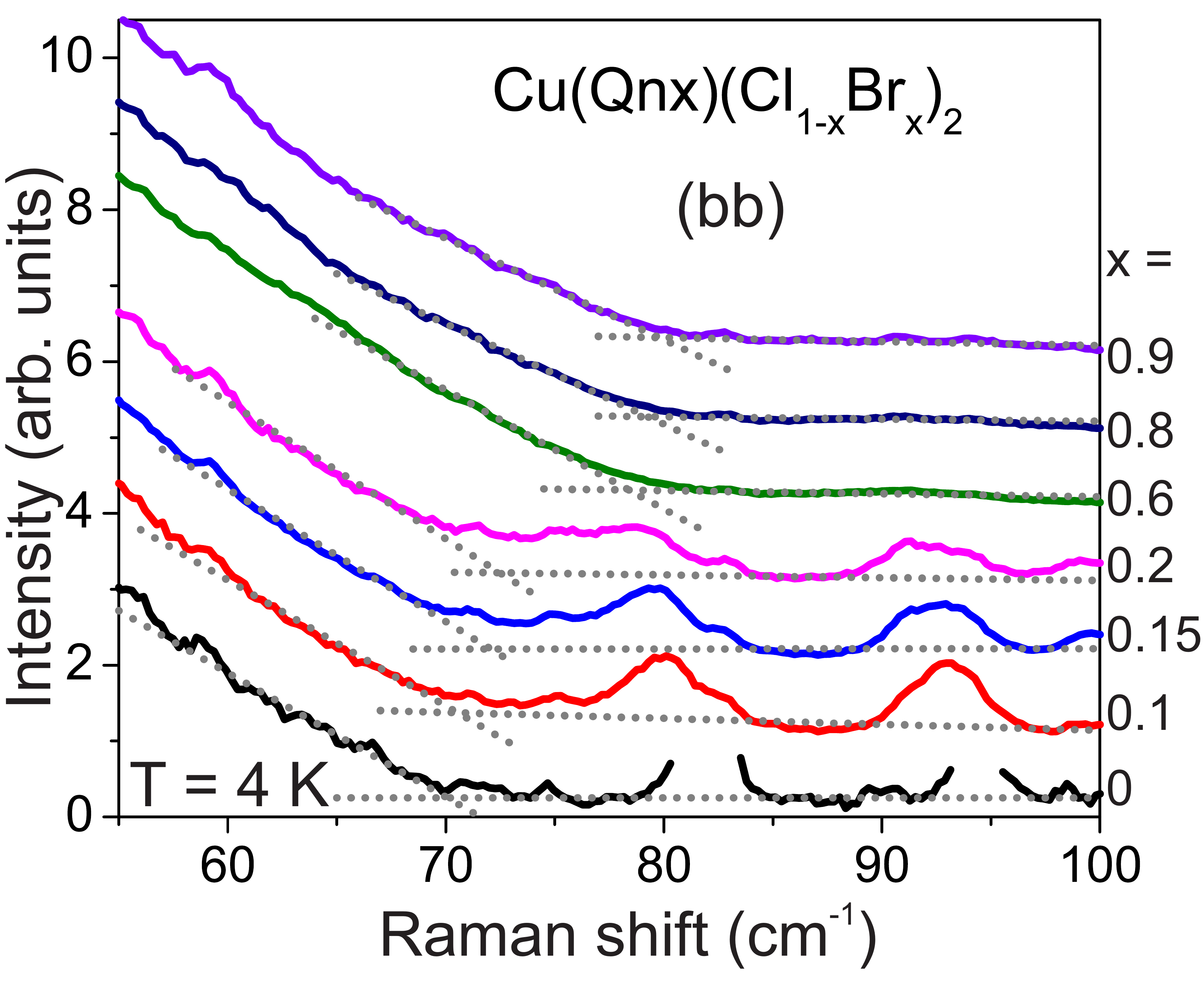}
\caption{The cutoff region for selected different concentrations. The spectra are offset for a clearer presentation of data. The fitted straight lines are presented as dotted lines showing the estimates of the cutoff. The sharp phonons for the x = 0 compound are truncated for the visual clarity of the data.\label{fig:cutoff}}
\end{figure}

In order to extract the information about the spin excitation gap and the magnon bandwidth, we studied the spectral region close to the onset $E_-$ (Fig.~\ref{fig:onset}) and upper bound $E_+$ (Fig.~\ref{fig:cutoff}) of the continuum. The corresponding threshold energies were determined empirically by linear extrapolation, as shown. We empirically fitted two straight lines below and above the thresholds, and defined the intersection of the two as the threshold value. The linear fit covered an energy range of $\pm 7$~cm$^{-1}$ around $E_-$, excluding the data in its immediate vicinity (0.6 cm$^{-1}$ on either side). A similar analysis was performed on the upper edge of the spectrum. In that case the linear fitting range was 15~cm$^{-1}$ below $E_+$ and up to 30~cm$^{-1}$ above it, excluding 1.2~cm$^{-1}$ in the immediate vicinity.  The obtained values for the upper and lower continuum thresholds are plotted in Figures \ref{fig:summary}a and \ref{fig:summary}b, respectively, in solid triangles.

\section{Discussion}
The overall shape of the observed scattering is qualitatively in very good agreement with theoretical predictions for a simple Heisenberg spin ladder with $J_\bot/J_{||}\sim1.7$. The two-magnon continuum is rather symmetric as expexted for present ratio of exchange constants. As Cl is replaced by Br, both the onset and the cutoff of the scattering increases steadily, suggesting that the energy scale involved in the exchange is increasing. More importantly, the cutoff increases at a greater rate than the onset. This indicates that upon increase of Br concentration, the exchange along the leg of the ladder is increasing faster than the exchange along the rung pushing the system towards the isotropic spin ladder case.

\begin{figure}
\centering
\includegraphics[width={\columnwidth}]{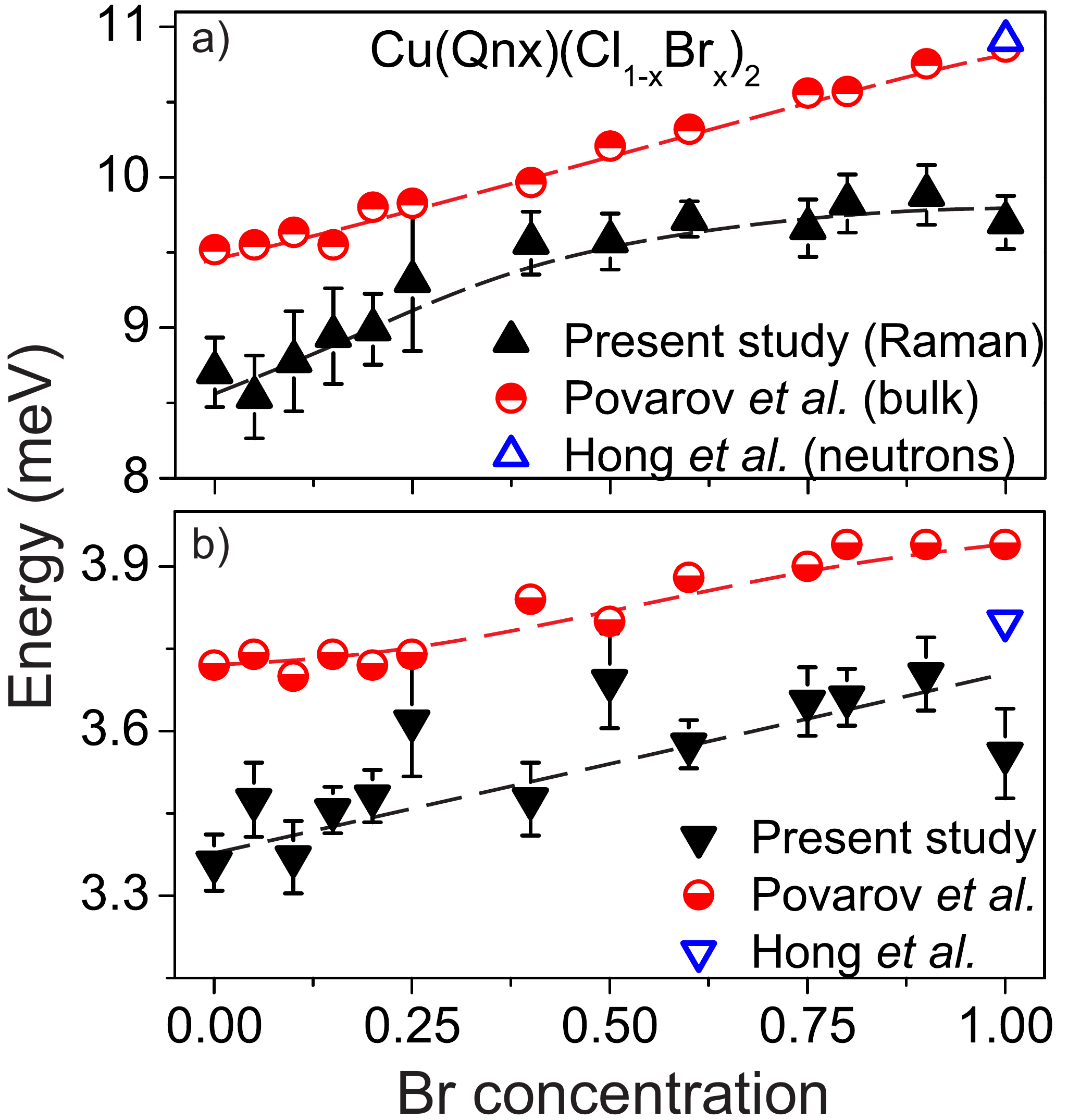}
\caption{Values of the onset(b) and cutoff(a) of magnetic scattering for different concentrations obtained in present study (black triangles). The estimates using the results from previous studies are also plotted. Red circles represent values estimated by using results of Povarov et al. and extrapolating DMRG calculations of Schmidiger et al. as described in the text. The blue triangles are taken from inelastic neutron study of Br-end compound by Hong {\it et al.}\cite{Hong2006} The dashed lines are guides to the eye. \label{fig:summary}}
\end{figure}

For a quantitative analysis we consider the extracted values for the onset and cutoff of the scattering continuum. It has been shown in \cite{Schmidt2001} that the onset of the scattering corresponds to the value of twice the gap size $\Delta$. We make use of these considerations for a comparison with expectations and related earlier studies. In Fig.~\ref{fig:summary}b, $2\Delta$ estimated from bulk magnetic measurements\cite{Povarov2014} and neutron scattering experiments\cite{Hong2006} is plotted in circles and open triangles respectively for a direct comparison with our result. Through the entire concentration range, the Raman data reproduces the trends observed by other techniques rather well. However, the measured threshold is consistently below estimates based on previous studies, by roughly $10~\%$.

Since Cu(Qnx)(Cl$_{1-x}$Br$_x$)$_2$ materials are strong-rung ladders, they are expected to have well-defined  single-magnon excitations across the entire Brillouin zone (see, for example, \cite{Schmidiger2013}). Assuming non-interacting magnons, we can relate $E_+$ to the upper bound of a $q=0$ 2-magnon continuum computed using the single-magnon dispersion relation. Computational studies have indeed shown this to be the appropriate description\cite{Schmidt2001} In this fashion we first estimated $E_+/J_{||}$ for two specific ladder species with $J_\bot/J_{||}=2$ and $J_\bot/J_{||}=1$, for which the magnon dispersion relation has been previously derived using DMRG techniques.\cite{Schmidiger2013} For the two cases, we obtain $E_+/J_{||}=6.28$  and $E_+/J_{||}=4.08$, respectively. We then used a linear interpolation of $E_+/J_{||}$ vs. $J_\bot/J_{||}$ to obtain  $E_+$ for the actual exchange constants in Cu(Qnx)(Cl$_{1-x}$Br$_x$)$_2$, as estimated by bulk magnetometry.\cite{Povarov2014} The resulting estimate for $E_+$ is plotted in semi-filled circles in Fig.~\ref{fig:summary}a. Once again, the experimental trend is well reproduced, but the observed upper Raman continuum threshold is typically 10\% below the estimate.

Bulk magnetic measurements may not provide a particualrly reliable estimate for $J_\bot/J_{||}$,  but actually are a very robust way to determine the spin gap $\Delta$. At least for the lower continuum threshold, the observed 10\% discrepancy therefore requires explanation. It seems that the most likely cause are terms in the spin Hamiltonian of CQX unaccounted for by the simple Heisenberg ladder model. In particular, the culprit could be weak inter-ladder coupling. If such interactions produced a 0.3~meV magnon dispersion perpendicular to the ladders axis, they would remain impossible to detect in the powder neutron scattering experiment of Ref.~\cite{Hong2006} On the other hand, the gap extracted from magnetometry assuming a purely one-dimensional model would not be correct. Compared to the true spin gap (the global minimum in the 3-dimensional magnon dispersion), it would be too large by about half the transverse magnon bandwidth, or 0.15~meV. That would account for the observed discrepancies on the lower threshold of the Raman continuum. The failings  of an analysis that assumes a purely one-dimensional model would then propagate to the estimate of $J_\bot/J_{||}$ and the upper continuum threshold.
 
\section{Conclusion}

In summary, we have obtained high quality magnetic raman scattering data on a family Cu(Qnx)(Cl$_{1-x}$Br$_x$)$_2$ which are close to being ideal Heisenberg systems. The observed scattering from two-magnon continuum has been found to shift to higher energies as more Br is introduced. The faster increase of the scattering sutoff is consistent with the system approaching the isotropic ladder. While the trend and observations are generally consistent with previous bulk measurements, some deviations persist. We hope that this work will stimulate numerical calculations of the exact shape of the Raman continuum for the partucular values of $J_\bot/J_{||}$ found in Cu(Qnx)(Cl$_{1-x}$Br$_x$)$_2$ for a direct comparison with experiment. Such a comparison will also clarify if the inconsistencies between gap energies deduced from Raman, neutron and bulk experiments are indeed due to inter-ladder coupling. Alternatively, the discrepancy may be due to an intrinsic feature of the Heisenberg ladder, such as magnon-magnon interactions.

\bibliography{CQXbib}

\end{document}